# Deep Learning of Structural Morphology Imaged by Scanning X-ray Diffraction Microscopy

## Authors

Aileen Luo[1], Tao Zhou[2], Martin V. Holt[2], Andrej Singer[1,*], Mathew J. Cherukara[3,*]

## Affiliations

[1]Department of Materials Science and Engineering, Cornell University, Ithaca, NY 14853, USA

[2]Center for Nanoscale Materials, Argonne National Laboratory, Lemont, IL 60439, USA

[3]Advanced Photon Source, Argonne National Laboratory, Lemont, IL 60439, USA

*mcherukara@anl.gov, asinger@cornell.edu

## Abstract

Scanning X-ray nanodiffraction microscopy is a powerful technique for spatially resolving nanoscale structural morphologies by diffraction contrast. One of the critical challenges in experimental nanodiffraction data analysis is posed by the convergence angle of nanoscale focusing optics which creates simultaneous dependency of the far-field scattering data on three independent components of the local strain tensor - corresponding to dilation and two potential rigid body rotations of the unit cell. All three components are in principle resolvable through a spatially mapped sample tilt series however traditional data analysis is computationally expensive and prone to artifacts. In this study, we implement NanobeamNN, a convolutional neural network specifically tailored to the analysis of scanning probe X-ray microscopy data. NanobeamNN learns lattice strain and rotation angles from simulated diffraction of a focused X-ray nanobeam by an epitaxial thin film and can directly make reasonable predictions on experimental data without the need for additional fine-tuning. We demonstrate that this approach represents a significant advancement in computational speed over conventional methods, as well as a potential improvement in accuracy over the current standard.

## Main Text

### Introduction

Scanning X-ray nanoprobe diffraction microscopy (SXDM) is a synchrotron radiation-based technique widely used to image the local distribution of nanoscale structural heterogeneities in materials. In particular, this non-destructive imaging method has provided numerous insights into the structure-properties relationships of battery nanoparticles[1, 2, 3], Mott insulators[4, 5], and ferroelectric materials[6, 7], among others, in a variety of sample environments or under operating conditions. This mode of imaging spatially resolves features down to 25 nm directly by diffraction contrast, thus bypassing the need for image reconstruction by solving the phase retrieval problem.

SXDM is performed by scanning a focused X-ray nanobeam across a sample and recording diffraction intensity on an area detector as a function of real space and reciprocal space coordinates, similar to 4D scanning transmission electron microscopy[8]. Images at a fixed Bragg angle can be collected rapidly, making the technique ideal for in-situ and operando measurements. Yet, real-time experimental feedback is limited to quantities in the detector frame, such as the total intensity and intensity centroids in pre-defined regions of interest (ROIs), rather than more directly relevant physical parameters in terms of the crystalline lattice. Specifically, lattice parameter changes cannot be disentangled from the lattice rotation in the scattering plane, as both affect the intensity distribution in the horizontal direction on the detector. Conversely, by rocking the crystal in Bragg diffraction geometry, 3D reciprocal space maps are measured at each real space coordinate. These maps are used to solve for the local lattice strain, rotation (tilt) of the reciprocal lattice vector within and out of the scattering plane[9]. Nevertheless, conventional rocking curve analysis requires extensive data collection time and careful image registration of multiple scans using a diffraction-invariant signal, which lends to often imprecise extractions of lattice information and a long feedback loop for informing subsequent measurements.

Recent developments in deep learning methods to analyze increasingly large imaging data have demonstrated great improvements in speed over conventional often computationally intensive methods. Broadly, applications requiring image reconstruction under noisy or sparse data conditions, such as medical imaging[10, 11] and electron microscopy[12, 13], benefit from the relative robustness against noise of machine learning approaches. With regards to coherent imaging techniques, these deep convolutional neural networks (CNNs) have been adapted to X-ray Bragg coherent diffractive imaging[14, 15, 16, 17, 18] and X-ray ptychography[19, 20], often with physically-aware training methods or architectures. Thus far, these implementations have focused largely on alternatives to phase retrieval for image reconstruction. While artificial intelligence-driven faster data analysis can also enable autonomous X-ray microscopy[21, 22], real-time analysis and feature recognition remains limited by incomplete diffraction contrast resolution.

In this study, we demonstrate a machine learning workflow that accurately identifies crystal lattice information from single-angle SXDM at speeds relevant to real-time analysis during an experiment. We introduce NanobeamNN, a light-weight CNN model that inputs images of Bragg diffraction intensity and outputs lattice strain and rotations. Our model is trained on simulated diffraction data from a highly crystalline epitaxial thin film that combines physical principles with the atypical shape of the convergent focused X-ray probe. The simulations require only knowledge of the beam, detector, and bulk crystal parameters. The X-ray energy, zone plate, and detector position are monitored instrument values, and the bulk sample can be characterized without the use of a synchrotron. This approach reduces the number of experimental measurements necessary to accurately understand local structural variations compared to conventional methods by enabling structural evaluation from 2D reciprocal space maps. We demonstrate that the neural network trained solely on simulated data can make predictions on experimental data that closely match and likely outperform state-of-the-art analysis. Additionally, the versatility of this framework and low computational requirements for training are potentially applicable to numerous systems of interest to the fields of materials science, solid-state physics, and chemistry.

## Results

### Model approach

The design of NanobeamNN uses a physics-aware simulated training dataset to target predictions of lattice strain and orientation. In the context of SXDM, which is primarily used to show contrast in structural morphology across a spatial region, one of the challenges of extracting the underlying crystal lattice information from reciprocal space lies in interpreting the effects of the focusing optics on the measured diffraction intensity. Hard X-ray scanning nanoprobes for dark-field microscopy typically use one of three types of focusing optics[23]: Kirkpatrick-Baez mirrors[24, 25], a multilayer Laue lens[26, 27], or a Fresnel zone plate[27, 28]. Here, we consider only the case of the zone plate, which focuses the beam diffractively, but we note that the approach can be applied to each. As shown in Figure 1a, the beam incident on the sample has a hollowed conical shape from the zone plate and central beam stop. An order sorting aperture selects only the first order of the diffractively focused beam, which is raster-scanned across the sample placed at the focal length. When in Bragg condition, highly ordered single crystal samples act as a mirror for the X-ray beam, reflecting the convergent beam shaped by focusing optics onto the pixel array detector. In such cases, the conventional analysis method of tracking shifts in intensity distribution is inherently limited because it does not consider the nature of the angular divergence of the diffracted beam and the intensity blocked by the central beam stop[29].

To incorporate the shape of the probe beam into the analysis, we first simulate diffraction of the X-ray nanoprobe by an epitaxially grown thin film. Specifically, we take a 2D projection of the Laue oscillations in the 3D reciprocal space construction (Methods, Simulated data generation) with added Poisson noise. The simulated diffraction patterns span a representative range of potential expected values for lattice strain $\epsilon$, tilt in-plane $\omega$, and tilt out-of-plane $\chi$, labeled following four-circle diffractometer notation. By visualizing the effects of varying these parameters on the resulting diffraction patterns (Supplementary Information, Fig. S1), it becomes apparent that strain changes the intensity distribution within the reflection of the beam (donut-shaped reflection), while lattice rotations change the position of the reflection. This understanding of measurable changes in the diffraction images informs our decision to choose a CNN architecture for NanobeamNN, which quantifies structural inhomogeneities in thin films measured by SXDM.

**Performance on simulated data**

Figure 1b illustrates the composition of NanobeamNN, which is similar in structure to other CNNs[18, 30]. Inputs are images of diffraction patterns, and outputs are predicted values that describe the crystal lattice. An activation function follows each convolutional layer, and a maximum pooling layer follows every other convolution, with one final fully connected layer. We label the simulated training data with known ground truth values used to generate the images, and NanobeamNN regressively learns to predict those parameters. Figure 2a shows the comparison between the diffraction patterns simulated with input parameters (strain, tilts) and the simulated diffraction generated by the strain and tilts predicted by NanobeamNN. The intensity distribution within the zone plate image (strain) and the position of the zone plate image within the detector frame (tilt) are correctly learned. The statistics of the network predictions for each parameter are detailed in the Supplementary Information, Figure S2. As shown in Figure 2b, the predicted diffraction closely resembles the experimentally measured data reported on $SrIrO_3$ thin films previously[31], particularly the position of the beam within the frame and the center of the intensity in the beam. This indicates that NanobeamNN correctly identifies the changes in diffraction corresponding to changes in the crystal lattice and demonstrates the versatility of the workflow.

The simulated images are scaled to a maximum of seven counts before the addition of noise (Poisson distribution) to match the signal level of the experimentally measured diffraction (Fig. 2b). To determine the effects of intensity level on model predictions, we evaluate the performance of networks trained on noise-free simulated data from 10 to 10,000 counts on noisy data of various intensity levels scaled to the training counts (Fig. S3). We find that a model trained on maximum diffraction intensity on the order of 10 photons per frame makes sufficiently accurate predictions on higher intensity data that is scaled to this range. Thus, it is advisable to collect experimental data on the order of 10 counts or more per diffraction pattern.

Subsequently, to test the model's performance in an imaging capacity, we simulate spatially distributed features similar to those that might be present in a sample. Figure 3a shows a 2D sinusoidal wave in strain and 2D sigmoidal functions in lattice rotations, which imitate experimentally observed domains and film lattice tilt due to substrate step edges, respectively. Accordingly, to discuss the comparison of NanobeamNN predictions (Fig. 3c) with the results from current analysis methods (Fig. 3b), we first introduce some underlying principles of conventional SXDM analysis and the relevant experimental constraints.

Somewhat intuitively, the only way to determine both the magnitude and direction of the momentum transfer vector Q in three dimensions is to measure scattering signal in a volume of reciprocal space[32]. Nonetheless, this measurement becomes less precise with a convergent beam due to the gradation in angles of the probe upon the sample and the blocking of the direct beam by the central stop. In a disordered sample, the hollowed ring of the zone plate optics would not be discernable in the measured diffraction[33, 34], and thus, the spread of scattered signal has no distinguishable correlation with the divergence angle of the beam. In such cases, one could track the shift in intensity distribution across multiple projections throughout the diffraction peak to establish trends in the local lattice heterogeneity[1, 5]. In a highly crystalline sample, however, the reflection of the nanobeam is clearly visible[7, 35] and the sample is more constrained (i.e. mechanically, such as in the case of a coherently strained epitaxial thin film on a single crystal substrate). It is precisely this highly crystalline constraint that allows us to model physically probably changes in the sample and their effects on 2D diffraction. The physics-aware simulation approach to training NanobeamNN circumvents the need to measure 3D reciprocal space maps (a slow process by scanning probe) and improves structural analysis by correcting for the shape of the beam.

Currently, this type of simulation is employed to infer strain and lattice rotation information without a full measurement of the 3D diffraction peak by fitting a weighted sum (interpolated center of mass) to find the highest correlation between each diffraction pattern and each simulated diffraction condition[36]. As shown in Figure 3b, we apply this method to fit simulated diffraction patterns generated using the values of the artificial sinusoidal and sigmoidal microscope images in Figure 3a. Even by fitting the correlation of simulated diffraction with our pre-generated library of training images that more than span the full range of strains and tilts represented, we are not able to fully separate features in strain and in-plane rotation. Furthermore, this method globally overestimates the magnitude of strain and can either over- or underestimate the in-plane rotation depending on the convoluting strain features. In comparison, NanobeamNN accurately resolves the feature trends in all parameters (Fig. 3c), more accurately predicts the magnitude of strain, and demonstrates a significant improvement in analysis time (see Discussion).

While the ground truth out-of-plane lattice rotation (Fig. 3a, right) is a smooth sigmoidal function, the results of both the correlation fitting method (Fig. 3b) and NanobeamNN predictions (Fig. 3c) show some discretization. These steps represent a quantification of the instrument resolution (resolution parameter $\tau$, defined in Methods, Simulated data generation) under this simulated experimental condition, which is influenced by the detector distance and pixel size, the X-ray energy, and the upsampling of the reciprocal space construction. We note that upsampling, a common technique in image processing, is used in simulation to improve the resolution of sub-pixel changes at the cost of computation time. The images are correspondingly downsampled to match the physical detector resolution at the final step of the simulation. The prediction values at which the steps in out-of-plane rotation occur correspond to the points at which the diffraction patterns shift by one pixel. The effects of the discrete detector pixels are not as pronounced for in-plane lattice rotation because the effects of strain and in-plane tilt on the intensity distribution in the horizontal direction on the detector are more nuanced.

**Performance on experimental data**

Although we demonstrate that NanobeamNN has the potential to provide more accurate quantification of lattice structure than conventional SXDM analysis methods through simulation, there remains a need for application to experimental data. Figure 4 shows a direct comparison between the analysis of an SXDM measurement by experiment-simulation correlation fitting (top row) and NanobeamNN (bottom row). The NanobeamNN results are the average predictions of ten models trained separately with different randomized model initializations and Poisson noise conditions. All models exhibit consistent metrics during training and validation, as well as consistent predictions on experimental data (Fig. S4), demonstrating robustness against noise. While we cannot comment definitively on which, if either, method is correct, we illustrate that the extracted feature maps are highly correlated (in particular, the strain maps have a Pearson correlation coefficient of ~0.89). Notably, the strain predictions by NanobeamNN are smaller in magnitude than those found via fitting, which aligns with the trend from simulated data shown in Figure 3. The relative improvement in SXDM data analysis of NanobeamNN is evident in consistency across both simulated and experimental data.

**Discussion**

The NanobeamNN workflow significantly reduces the computational time needed to extract structural parameters from SXDM data. The experimental image shown in Figure 4 has 165 x 165 points; calculating the correlation matrix for and fitting three parameters from 64 x 64 pixel diffraction patterns takes approximately 10 hours using close to 1 TB of RAM. The computation time scales linearly with the number of data points and depends on the amount of available memory. Conversely, NanobeamNN makes predictions on the same data within seconds, a factor of around 1000 improvement. Moreover, once the neural network is trained, it can directly make predictions on any measurement carried out at the same X-ray energy, Bragg condition, film thickness, and detector parameters, without the need for adjustments, further reducing computational costs across multiple SXDM scans. We note that although generation of the expansive simulated dataset is somewhat computationally costly (see Methods, Simulated data generation), it is highly parallelizable and completed before an experiment, as well as being a necessary component of both the conventional correlation fitting analysis and NanobeamNN training. The time required to train the neural network and make predictions will vary depending on one's machine; however, the network is small enough to be trained on a personal computer without the use of GPUs in a reasonable timeframe (see Methods, Neural network architecture and training).

With further development, NanobeamNN can be adapted for real-time analysis of lattice heterogeneity and generalizability to more tunable instrument conditions, such as consideration of X-ray energy in resonant diffraction experiments. This implementation is designed to be trained once for each SXDM experiment and requires that the count rate be held at the same order of magnitude as that used in the training data for the duration of the experiment. Notably, due to the physical basis of Laue oscillations for the scattering model, these analyses (both correlation fitting and NanobeamNN) methods are limited to films of exceptional quality in which these oscillations can be observed. Consequently, we estimate that the optimal thickness range is between 5 nm and 300 nm, with loss of strain (intensity distribution throughout the zone plate reflection) resolution approaching the lower bound and loss of thickness fringe resolution approaching the upper bound. The exact constraints will also depend on the diffraction resolution parameter $\tau$ (Methods, Simulated data generation). Despite these restrictions on the sample choice, we emphasize that many of the functional materials characterized by SXDM meet the criteria for NanobeamNN prediction.

In conclusion, NanobeamNN provides several critical improvements to SXDM data analysis. It significantly reduces the analysis time by using physical principles to model nanobeam diffraction, as well as increases the accuracy of lattice strain and orientation quantification over conventional methods in cases of highly crystalline epitaxial thin film samples. Current real-time experimental feedback consists of the total intensity within in a pre-defined ROI, as well as the shift in intensity distribution across detector coordinates (Fig. S5), which does not resolve lattice strain from in-plane rotation. NanobeamNN, with its physics-aware pre-training method using known instrument, bulk crystal, and Bragg diffraction conditions, enables quantification of the lattice structural morphology at a rate applicable to real-time analysis during experimental data acquisition with an edge computing device[22].

## Methods

**Neural network architecture and training**

The NanobeamNN model is a convolutional neural network that predicts crystalline lattice information from x-ray nanoprobe diffraction. As shown in Fig. 1b, the model trains only on inputs of simulated focused nanobeam diffraction patterns, and outputs values of strain and lattice rotation about two axes. We use a PyTorch implementation of 2D convolution layers and maximum pooling, followed by a fully connected layer, with a mean squared error (squared L2 norm) loss criterion to regressively learn these physically relevant parameters.

The training dataset of simulated diffraction patterns spans ranges of strain and lattice rotation values that cover structural heterogeneities potentially observed in experiments. As described in the section "Simulated data generation", 2D diffraction of an X-ray nanobeam from a single crystal thin film was modeled using the physics principle of Laue oscillations. The entire simulated dataset consists of 68,921 images of 64 x 64 pixels each, which is randomly split into training (80%), validation (10%), and test (10%) sets. To ensure model robustness, we add representative noise to each image according to the Poisson distribution. The training cycle uses the adaptive moment estimation (ADAM) optimizer with a learning rate of 0.0001 to update the weights and biases of the model. Each epoch also includes a validation cycle to evaluate the performance of the network. The network trains on a single NVIDIA GeForce RTX 3090 GPU for 500 epochs, which takes ~40 minutes with a batch size of 64.

**Simulated data generation**

Diffraction of a zone plate optics focused X-ray nanoprobe (convergent beam) by a highly crystalline epitaxially grown thin film is modeled using geometric optics by taking a 2D projection of the Laue oscillations in the 3D reciprocal space construction from known bulk sample and instrument parameters. We generate simulated images of a sinc function convoluted with the position of the diffracted beam on the detector, given by the relative position and orientation of the crystal to the focusing optics and detector. The method is described in the equation below:

$$I(N, c, l, \epsilon, \omega, \chi, d_{CS}, d_{ZP}, E, R, \delta) = \text{conv}\big(L(N, c, l, \epsilon, \omega, \chi), O(d_{CS}, d_{ZP})\big) \cdot D(E, c, l, R, \delta);$$

where $N$ is the number of unit cell layers in the film (as measured by *in situ* reflection high energy electron diffraction or X-ray reflectivity), $c$ is the lattice constant in Angstroms (as measured by bulk X-ray diffraction), and $\epsilon = \frac{\Delta d_{00l}}{d_{00l}}$ is the lattice strain in the $[00\,l]$ direction where $\epsilon \in$

$[−0.005, 0.005]$, corresponding to the $00l$ diffraction peak. We denote the lattice rotation (tilt) angle in the scattering plane as $\omega$ and out of the scattering plane as $\chi$, where $\omega \in [−0.05°, 0.05°]$ and $\chi \in [−0.1°, 0.1°]$. The reciprocal space construction is given by the function L. O represents the masking of scattered intensity by the angles of the beam, as determined by the diameters of the central beam stop and the zone plate, $d_{CS}$ and $d_{ZP}$, respectively. Finally, the detector function D, which defines the slice of the Ewald sphere being measured depends on the X-ray energy (keV) $E$, $c$, $l$, sample-detector distance $R$ in meters, and the detector pixel size $\delta$ in meters. The detector resolution for any given experimental setup is limited by the following parameter $\tau$:

$$\tau(E, R, \delta) = \frac{\delta}{R} \cdot \frac{2\pi \cdot E}{12.398 \text{ keV} \cdot \text{Å}}$$

We note that upsampling of the reciprocal space coordinates is used in simulation to improve the resolution of sub-pixel changes near edge features at the cost of computation time. The final detector images are downsampled to match the physical detector size.

**Data acquisition**

The experimental methods, including sample synthesis and scanning X-ray nano-probe data acquisition have been reported previously[31]. Synchrotron light sources have traditionally provided unique insight into condensed matter physics systems due both to their high penetrating power and the fundamental monochromaticity of the X-ray beam (typically 10^-4 or greater at third generation sources, e.g. a 1eV bandpass on a 10keV beam). This gives the potential for studying complex phenomena within extended volumes at a strain resolution (dc/c ~10^-4 = 0.01%) that is several orders of magnitude higher than achievable with conventional electron microscopy. Especially with the advent of diffraction limited storage rings this sensitivity is now being harnessed for real-space imaging of crystalline ordering through nano-focusing optics however a key challenge in this methodology is the beam convergence angle introduced by the focusing optics which convolutes micro-radian level lattice curvatures with lattice strain in the far-field scattering pattern.

**Experimental data analysis**

Experimental data is used to test the performance of NanobeamNN against state-of-the-art conventional data analysis methods. A region of interest (ROI) of 128 x 128 detector pixels is selected to encompass the entire measured diffraction peak. The pixels of this ROI are binned by aggregating each 2 x 2 array into one value, for a final image size of 64 x 64 pixels, to improve the signal to noise ratio. Currently, the only method to infer strain and lattice rotation information without a full measurement of the 3D diffraction peak is by fitting a weighted sum (center of mass interpolation) to find the highest correlation between each diffraction pattern and each simulated diffraction condition.

## Data availability

The simulated and experimental datasets that support the findings of this study will be made available before publication once approval has been obtained from our instituion.

## Code availability

The code and trained model developed in this study will be made available before publication in a public GitHub repository once approval has been obtained from our institution.

## References

1. Li, J. *et al.* Probing lattice defects in crystalline battery cathode using hard X-ray nanoprobe with data-driven modeling. *Energy Storage Mater.* **45**, 647–655 (2022).

2. Judge, W. J. *et al.* Evaluation of Chemical and Structural Homogeneity in Single Particles of $Li_{1-x}Ni_{0.33}Mn_{0.33}Co_{0.33}O_2$. *J. Phys. Chem. C* **126**, 16082–16089 (2022).

3. Liu, T. *et al.* Rational design of mechanically robust Ni-rich cathode materials via concentration gradient strategy. *Nat. Commun.* **12**, 6024 (2021).

4. Shabalin, A. G. *et al.* Nanoscale Imaging and Control of Volatile and Non-Volatile Resistive Switching in $VO_2$. *Small* **16**, 2005439 (2020).

5. Singer, A. *et al.* Nonequilibrium Phase Precursors during a Photoexcited Insulator-to-Metal Transition in $V_2O_3$. *Phys. Rev. Lett.* **120**, 207601 (2018).

6. Lummen, T. T. A. *et al.* Thermotropic phase boundaries in classic ferroelectrics. *Nat. Commun.* **5**, 3172 (2014).

7. Guzelturk, B. *et al.* Sub-Nanosecond Reconfiguration of Ferroelectric Domains in Bismuth Ferrite. *Adv. Mater.* **35**, 2306029 (2023).

8. Ophus, C. Four-Dimensional Scanning Transmission Electron Microscopy (4D-STEM): From Scanning Nanodiffraction to Ptychography and Beyond. *Microsc. Microanal.* **25**, 563–582 (2019).

9. Holt, M., Harder, R., Winarski, R. & Rose, V. Nanoscale Hard X-Ray Microscopy Methods for Materials Studies. *Annu. Rev. Mater. Res.* **43**, 183–211 (2013).

10. Zhu, B., Liu, J. Z., Cauley, S. F., Rosen, B. R. & Rosen, M. S. Image reconstruction by domain-transform manifold learning. *Nature* **555**, 487–492 (2018).

11. Musleh, A. Computed Tomography (Ct) Scan Assisted Machine Learning in the Management of Artifacts Related to Paranasal Sinuses and Anterior Cranial Fossa. *Comput. Intell. Neurosci.* **2022**, 1–9 (2022).

12. Zhou, T., Cherukara, M. & Phatak, C. Differential programming enabled functional imaging with Lorentz transmission electron microscopy. *Npj Comput. Mater.* **7**, 141 (2021).

13. Ziatdinov, M., Ghosh, A., Wong, C. Y. & Kalinin, S. V. AtomAI framework for deep learning analysis of image and spectroscopy data in electron and scanning probe microscopy. *Nat. Mach. Intell.* **4**, 1101–1112 (2022).

14. Cherukara, M. J., Nashed, Y. S. G. & Harder, R. J. Real-time coherent diffraction inversion using deep generative networks. *Sci. Rep.* **8**, 16520 (2018).

15. Wu, L., Juhas, P., Yoo, S. & Robinson, I. Complex imaging of phase domains by deep neural networks. *IUCrJ* **8**, 12–21 (2021).

16. Lim, B. *et al.* A convolutional neural network for defect classification in Bragg coherent X-ray diffraction. *Npj Comput. Mater.* **7**, 115 (2021).

17. Chan, H. *et al.* Rapid 3D nanoscale coherent imaging via physics-aware deep learning. *Appl. Phys. Rev.* **8**, 021407 (2021).

18. Yao, Y. *et al.* AutoPhaseNN: unsupervised physics-aware deep learning of 3D nanoscale Bragg coherent diffraction imaging. *Npj Comput. Mater.* **8**, 124 (2022).

19. Cherukara, M. J. *et al.* AI-enabled high-resolution scanning coherent diffraction imaging. *Appl. Phys. Lett.* **117**, 044103 (2020).

20. Hoidn, O., Mishra, A. A. & Mehta, A. Physics constrained unsupervised deep learning for rapid, high resolution scanning coherent diffraction reconstruction. *Sci. Rep.* **13**, 22789 (2023).

21. Kandel, S. *et al.* Demonstration of an AI-driven workflow for autonomous high-resolution scanning microscopy. *Nat. Commun.* **14**, 5501 (2023).

22. Babu, A. V. *et al.* Deep learning at the edge enables real-time streaming ptychographic imaging. *Nat. Commun.* **14**, 7059 (2023).

23. Sakdinawat, A. & Attwood, D. Nanoscale X-ray imaging. *Nat. Photonics* **4**, 840–848 (2010).

24. Mimura, H. *et al.* Hard X-ray Diffraction-Limited Nanofocusing with Kirkpatrick-Baez Mirrors. *Jpn. J. Appl. Phys.* **44**, L539 (2005).

25. Osterhoff, M. *et al.* Focus characterization of the NanoMAX Kirkpatrick–Baez mirror system. *J. Synchrotron Radiat.* **26**, 1173–1180 (2019).

26. Kang, H. C. *et al.* Focusing of hard x-rays to 16 nanometers with a multilayer Laue lens. *Appl. Phys. Lett.* **92**, 221114 (2008).

27. Nazaretski, E. *et al.* Design and performance of an X-ray scanning microscope at the Hard X-ray Nanoprobe beamline of NSLS-II. *J. Synchrotron Radiat.* **24**, 1113–1119 (2017).

28. Winarski, R. P. *et al.* A hard X-ray nanoprobe beamline for nanoscale microscopy. *J. Synchrotron Radiat.* **19**, 1056–1060 (2012).

29. Ying, A. *et al.* Modeling of kinematic diffraction from a thin silicon film illuminated by a coherent, focused X-ray nanobeam. *J. Appl. Crystallogr.* **43**, 587–595 (2010).

30. Simonyan, K. & Zisserman, A. Very Deep Convolutional Networks for Large-Scale Image Recognition. (2014) doi:10.48550/ARXIV.1409.1556.

31. Luo, A. *et al.* X-ray nano-imaging of defects in thin film catalysts via cluster analysis. *Appl. Phys. Lett.* **121**, 153904 (2022).

32. Warren, B. E. *X-Ray Diffraction*. (Dover Publications, Inc, New York, 1990).

33. Chen, Y. *et al.* Seeded Lateral Solid-Phase Crystallization of the Perovskite Oxide SrTiO$_3$. *J. Phys. Chem. C* **123**, 7447–7456 (2019).

34. Calvo-Almazan, I. *et al.* Strain Mapping of CdTe Grains in Photovoltaic Devices. *IEEE J. Photovolt.* **9**, 1790–1799 (2019).

35. Li, X., Luo, Y., Holt, M. V., Cai, Z. & Fenning, D. P. Residual Nanoscale Strain in Cesium Lead Bromide Perovskite Reduces Stability and Shifts Local Luminescence. *Chem. Mater.* **31**, 2778–2785 (2019).

36. Roy, P. *et al.* Origin of Topological Hall-Like Feature in Epitaxial SrRuO$_3$ Thin Films. *Adv. Electron. Mater.* **9**, 2300020 (2023).

## Acknowledgements

This material is based upon work supported by the U.S. Department of Energy, Office of Science, Office of Workforce Development for Teachers and Scientists, Office of Science Graduate Student Research (SCGSR) program. The SCGSR program is administered by the Oak Ridge Institute for Science and Education for the DOE under contract number DE-SC0014664. A.S. acknowledges the support by the U.S. Department of Energy, Office of Science, Office of Basic Energy Sciences (Contract No. DE-SC0019414). M.J.C also acknowledge support from the U.S. Department of Energy, Office of Science, Office of Basic Energy Sciences Data, Artificial Intelligence, and Machine Learning at DOE Scientific User Facilities program under Award Number 34532. This research used resources of the Advanced Photon Source and Center for Nanoscale Materials, both U.S. Department of Energy (DOE) Office of Science user facilities at Argonne National Laboratory and is based on research supported by the U.S. DOE Office of Science-Basic Energy Sciences, under Contract No. DE-AC02-06CH11357.

## Author contributions

A.L. adapted the simulations to experimental conditions, built the neural network model, and performed network training and testing, with advice from T.Z., A.S., and M.J.C. T.Z. and M.V.H.

developed the simulation and correlation fitting method. A.L., T.Z. A.S., and M.J.C. contributed the proposal of the initial idea and the manuscript writing.

## Competing interests

The authors declare no competing interests.

# Figures

**Figure 1: Schematic illustration of experimental setup and neural network architecture.**

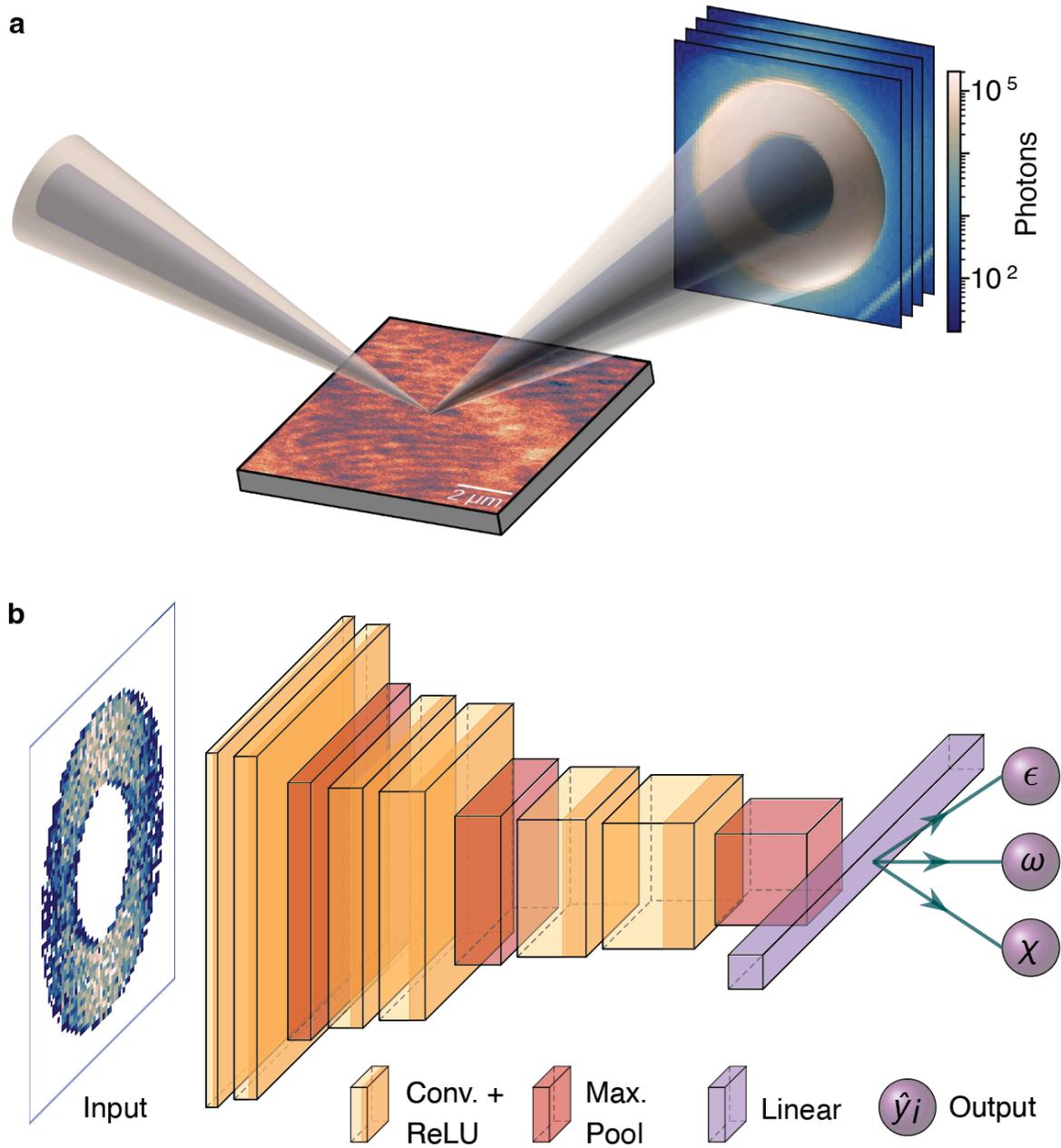

**a** Configuration of an archetypal scanning X-ray nanoprobe diffraction microscopy experiment in Bragg geometry. The hollowed cone on the left represents the X-ray beam focused onto the sample using zone plate optics. The diffracted intensity is measured on a 2D pixel array detector.

**b** Diagram of the layers in the neural network (box dimensions not drawn to scale). Inputs are simulated diffraction patterns and outputs are strain and lattice rotation values.

**Figure 2: NanobeamNN predictions on simulated and experimental data.**

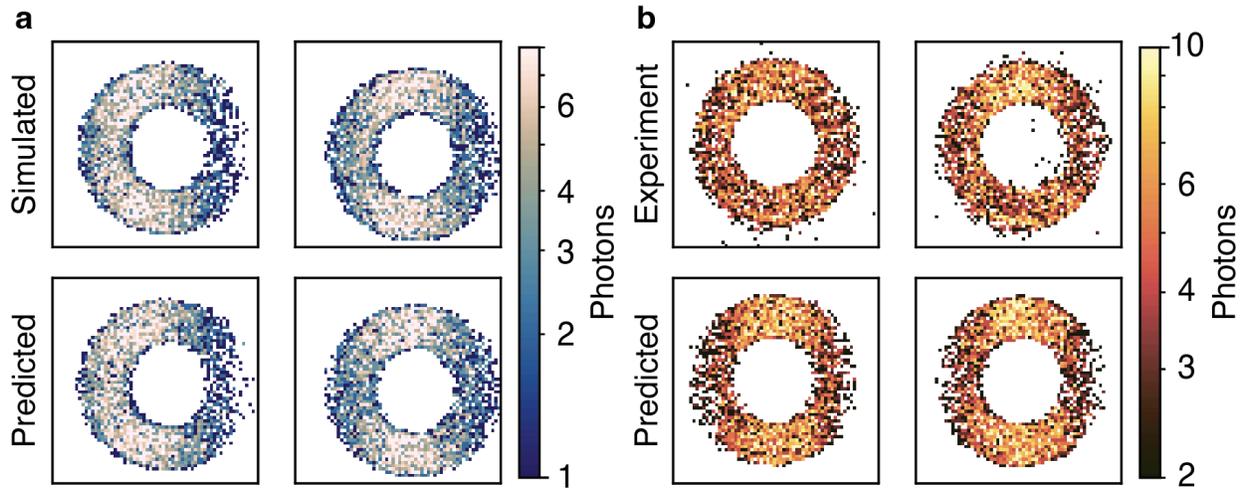

**a** Top row: simulated diffraction patterns directly input to NanobeamNN. Bottom row: simulated diffraction generated from the values of strain $\epsilon$ and tilts $\omega$ and $\chi$, predicted by NanobeamNN from the images in the top row. **b** Top row: experimentally measured diffraction patterns of the pseudo-cubic (pc) $002_{pc}$ peak of an $SrIrO_3$ thin film. Bottom row: simulated diffraction generated from predicted $\epsilon$, $\omega$, and $\chi$, from the images in the top row.

**Figure 3: Feature extraction from simulated scanning X-ray nanoprobe diffraction microscopy measurement by conventional fitting and NanobeamNN.**

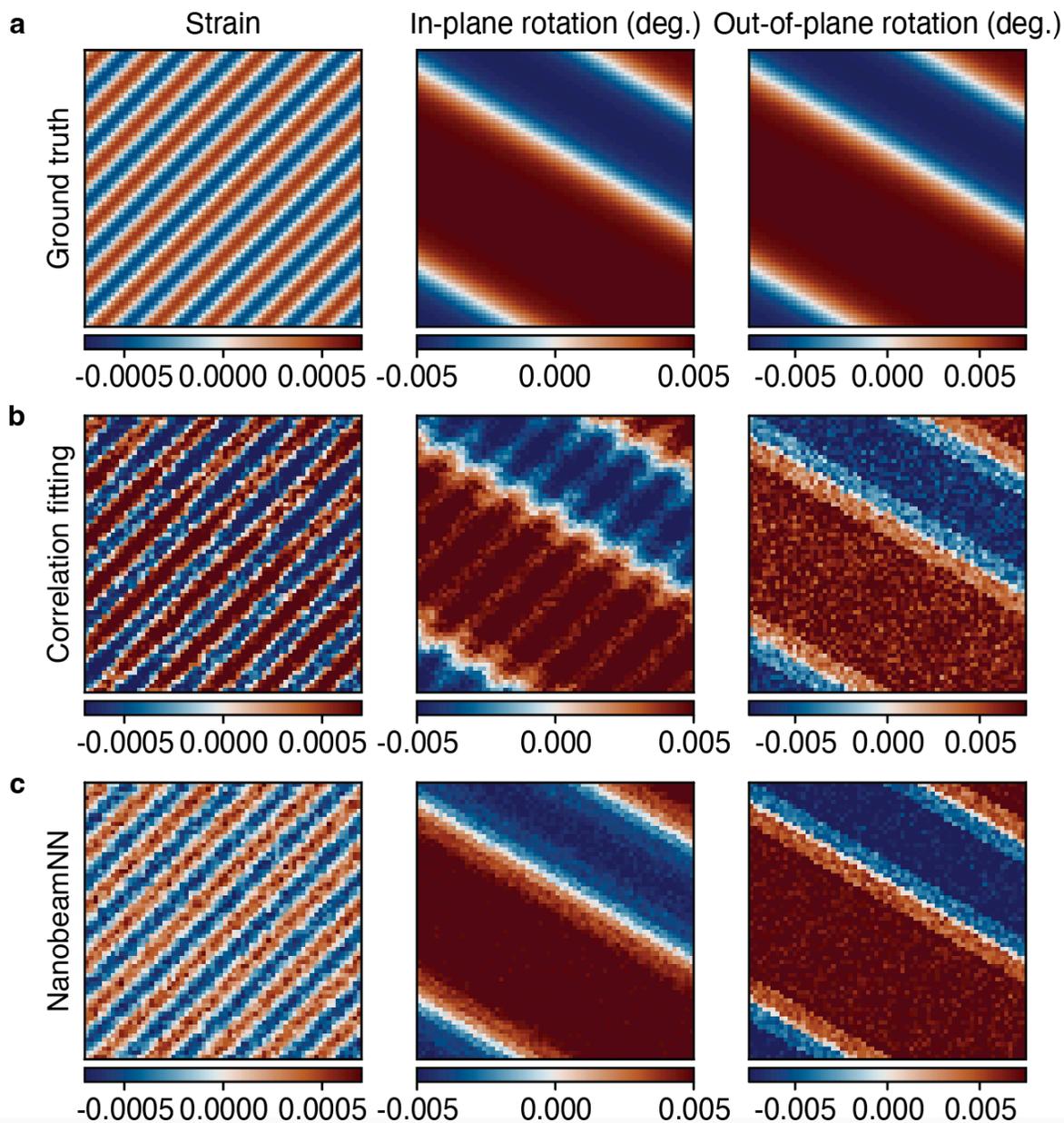

**a** Simulated spatial distribution of strains $\epsilon(x, y)$ and tilts $\omega(x, y)$, and $\chi(x, y)$, which mimic spatially distributed features present in a scanning X-ray nanoprobe diffraction microscopy measurement. **b** $\epsilon(x, y)$, $\omega(x, y)$, and $\chi(x, y)$, as analyzed by conventional fitting of the correlation between measured and simulated diffraction. **c** $\epsilon(x, y)$, $\omega(x, y)$, and $\chi(x, y)$, predicted by NanobeamNN.

**Figure 4: Comparison of conventional fitting and NanobeamNN on experimental data.**

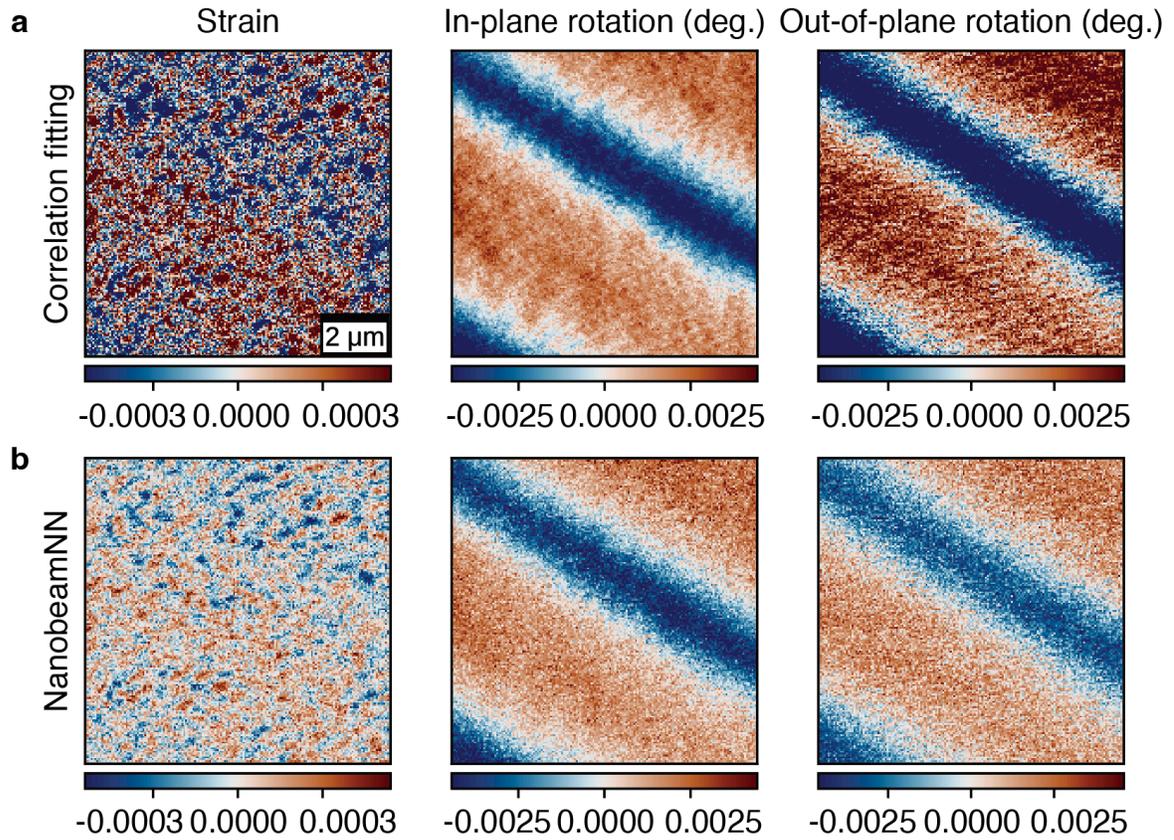

**a** Diffraction contrast features of the SrIrO$_3$ 002$_{pc}$ peak measured by scanning X-ray nanoprobe diffraction microscopy as analyzed by conventional fitting of the correlation between measured and simulated diffraction. **b** NanobeamNN evaluations of the experimental data shown in (**a**).